\documentclass[twocolumn,showpacs,preprintnumbers,prb]{revtex4}
\usepackage{dcolumn}
\usepackage{amssymb}
\usepackage{bm}
\usepackage{graphicx}
\usepackage{float}
\usepackage{color}
\usepackage{epstopdf}
\newcommand{\avg}[1]{\left< #1 \right>} 
\newcommand{\ket}[1]{| #1 \rangle} 
\newcommand{\bra}[1]{\langle #1 |} 

\begin{document}

\title{Entanglement Area Law in Disordered Free Fermion Anderson Model in One, Two, and Three Dimensions
}
\author{Mohammad Pouranvari, Yuhui Zhang and Kun Yang}
\affiliation{National High Magnetic Field Laboratory and Department of Physics,
Florida State University, Tallahassee, Florida 32306, USA}

\pacs{}
\date{\today}
\begin{abstract}

We calculate numerically the entanglement entropy of free fermion ground states in one-, two- and three-dimensional Anderson models, and find that it obeys the area law as long as the linear size of the subsystem is sufficiently larger than the mean free path. This result holds in the metallic phase of the three-dimensional Anderson model, where the mean free path is finite although the localization length is infinite. Relation between the present results and earlier ones on area law violation in special one-dimensional models that support metallic phases is discussed.

\end{abstract}

\maketitle

\section{Introduction}
Recent years have witnessed tremendous progress in the study of entanglement in condensed matter/many-body physics. Among these studies, free fermion systems play a
very special role.\cite{peschel} Simple as they may seem, fermions are intrinsically non-local, due to the {\em anti-}commutation relation fermion operators satisfy,
no matter how far apart they are. Such non-locality shows up as {\em enhanced} entanglement in the ground state; for example for many years Fermi sea states were the
only known ground states whose block entanglement entropy (EE) violates the area law satisfied by most ground states above 1D.\cite{GioevKlich,wolf,swingle} It is only recently
shown that a similar violation occurs in interacting fermion systems in the Fermi liquid phase,\cite{dingprx12} and bosonic models with excitation spectra that vanish
on (extended) Bose surfaces.\cite{LaiYangBonesteel} The existence of sharp Fermi or Bose surfaces is crucial for the area law violation in translationally invariant systems.

Comparatively much less effort has been devoted to studies of fermions in the presence of disorder potential. In a recent work\cite{PouranvariYang14} we studied two very
special (one-dimensional) 1D models that exhibit free fermion metal-insulator transition (MIT), and found area law violation in the metallic phase, {\em despite} the presence of
disorder, and thus absence of sharp Fermi surface (actually points in 1D). It was conjectured\cite{PouranvariYang14} that as long as the system is metallic, namely
states are de-localized at the Fermi energy, there will be area-law violation. In the present work we test this conjecture by performing detailed numerical studies of the Anderson model\cite{anderson} in one-, two and three-dimensions. We find that the area law is actually respected in all cases, including the metallic phase in 3D. We do observe an enhancement (beyond area law) as systems sizes increase while below the mean free path; such enhancement disappears once the system size becomes sufficiently bigger than the mean free path. The origin of the difference between the Anderson model studied here and the special models studied earlier\cite{PouranvariYang14} will be discussed.

The remainder of the paper is organized as follows. In sec. II we introduce our model and numerical method for calculating EE. Results of our calculations are presented in sec. III. Sec. IV offers a summary and discussions on our results.

\section{Model and Basic Considerations}
Anderson model in $D$ dimension is a model with constant nearest neighbour hopping term and random on-site energy $\varepsilon$:

\begin{equation}
H=\sum _{\vec{r}} \sum_{\vec{d}} (c_{\vec{r}} ^{\dagger} c_{\vec{r}+\vec{d}}+ c_{\vec{r}+\vec{d}} ^{\dagger} c_{\vec{r}}) +
\sum _{\vec{r}} \varepsilon_{\vec{r}} c_{\vec{r}} ^{\dagger} c_{\vec{r}},
\end{equation}
where summation is over all sites in $D$ dimensional hyper cubic lattice (with lattice constant set to be 1) and $\vec{d}$ is a vector connecting a site to its nearest neighbour. $\varepsilon$'s are uniformly distributed between $-w/2$ and $w/2$. \cite{Laflorencie} The Fermi energy $E_F$ is set to be 0 (so the lattice is half-filled) in all cases, while in 2D we also study $E_F=1$ to avoid the van Hove singularity at the band center. We consider cubic-shaped finite-size systems with linear size $L$ and open-boundary conditions. We then divide them into two equal subsystems $A$ and $B$ with size $L^{D-1}\times (L/2)$, and calculate the disorder-averaged entanglement entropy as detailed below.

For a system in a pure state $\ket{\psi}$, the density matrix is $\rho= \ket{\psi} \bra{\psi}$. Reduced density matrix of each subsystem (A or B) is obtained by tracing over degrees of freedom of the other subsystem: $\rho^{A/B}=tr_{B/A} (\rho)$. Block EE between the two subsystems is $EE=-tr(\rho^{A}\ln{\rho^{A}})=-tr(\rho^{B}\ln{\rho^{B}})$. For a single Slater-determinant ground state,
\begin{equation} \label{rho}
\rho^{A/B}=\frac{1}{Z} e^{-H^{A/B}}
\label{entanglementH}
\end{equation}
are characterized by free-fermion {\em entanglement} Hamiltonians
\begin{equation}
H^{A/B} = \sum_{ij} h_{ij}^{A/B}  c_{i} ^{\dagger} c_{j},
\end{equation}
where $Z$ is determined by the normalization condition $tr \rho^{A/B}=1$. We calculate EE using method of Ref. \onlinecite{correl} by diagonalizing correlation matrix of subsystem A
\begin{equation}
C_{mn}=\avg{c_{m} ^{\dagger} c_{n}},
\end{equation}
and find its eigenvalues $\zeta$'s. Then EE takes the form
\begin{equation}
\text{EE}=-\sum_{l=1} ^{N_A} [\zeta_l \ln(\zeta_l)+(1-\zeta_l) \ln(1-\zeta_l)],
\end{equation}
where $N_A$ is number of sites in subsystem $A$.

In one and two dimensions, all states are localized with any finite disorder. However there is an important difference between them: In 1D the localization length $\xi$ is of the same order as mean free path $\ell$, while in 2D we have $\xi\gg\ell$ for weak disorder. In 3D there is a metal-insulator transition (MIT) at a critical value of disorder strength $w_c \approx 16$, \cite{MacKinnonKramer_w} where $\xi$ diversges.
The focus of our numerical calculation is the interplay of the three different length scales, mean free path $\ell$ (calculated perturbatively in the Appendix), localization length $\xi$, and (sub)system size $L$, and their effects on entanglement.

\section{Results}

\subsection{Anderson Model in One and Two Dimensions}

In these two cases, all states are localized as long as $w > 0$.

\begin{figure}
\includegraphics[scale=0.2]{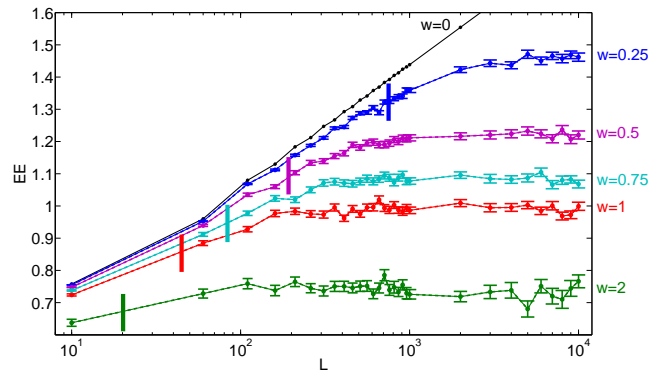}
\caption{\label{1danderson}[Color online] Entanglement Entropy of one dimensional Anderson model in log-linear scale for different value of disorder strength $w$. $E_F=0$. Mean free path corresponding to selected $w$'s is indicated as a vertical line. Horizontal axis is the linear size of the system, $L$, varying from $10$ up to $10000$. Number of samples at each point of $L$ is $200$. The strange break in the small $w$'s including $w=0$ case is a consequence of small size effect in calculating average EE. We see that for bigger sizes the behavior is more smooth.}
\end{figure}

Fig. \ref{1danderson} shows 1D EE as a function of system size $L$ for different values of $w$. As size $L$ increases, EE
grows logarithmically for $w=0$ as expected. For $w > 0$, EE grows with $L$ in a manner similar to the disorder free case up to some point, and then saturates, indicating area law is obeyed for sufficiently large system sizes. We find substantial deviation (from $w=0$ case) starts when the system size $L$ reaches the mean free path $\ell$, and saturation occurs around $L\approx 3\ell$. We note in 1D we have the localization length $\xi\sim \ell$; it is thus not immediately clear at this point which of the two controls the crossover.

\begin{figure}
\includegraphics[scale=0.2]{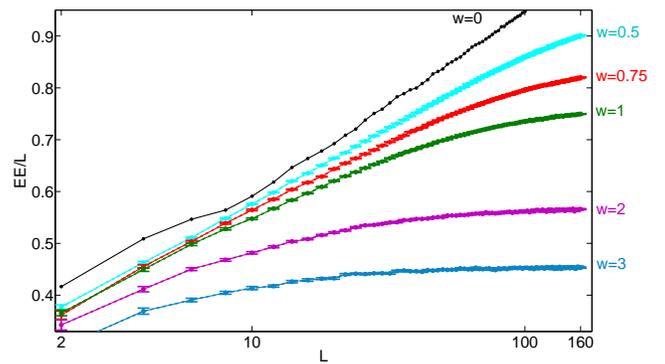}
\caption{\label{2danderson0}[Color online] Entanglement Entropy divided by linear system size of two dimensional Anderson model in log-linear scale for different value of disorder strength $w$. $E_F=0$. Horizontal axis is the linear size of the system, $L$. The total system has $L \times L$ sites. Number of samples at each point of $L$ is $100$.}
\end{figure}

\begin{figure}
\includegraphics[scale=0.2]{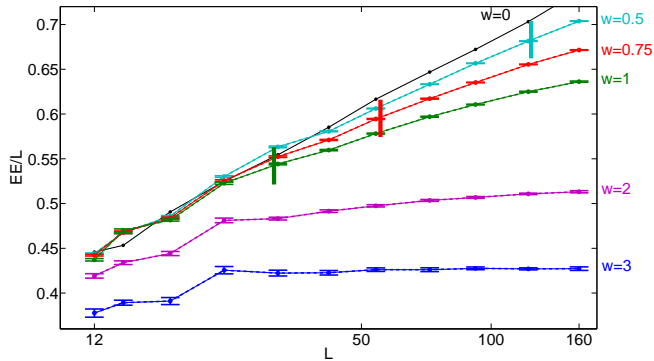}
\caption{\label{2danderson1}[Color online] Entanglement Entropy divided by linear system size of two dimensional Anderson model in log-linear scale for different value of disorder strength $w$. $E_F=1$. Mean free path corresponding to selected $w$'s is indicated as a vertical line. Horizontal axis is the linear size of the system, $L$. The total system has $L \times L$ sites. Number of samples at each point of $L$ is $100$.}
\end{figure}

In Figs. \ref{2danderson0} and  \ref{2danderson1}, 2D EE divided by boundary length $L$ (to account for the area law contribution) as a function of $L$ for different values of disorder strength is plotted for $E_F=0$ and $E_F=1$ respectively. We see again that, for $w=0$ there is a logarithmic growth as expected,\cite{GioevKlich,wolf} while for $w > 0$ such growth stops beyond certain length scale, indicating area law behavior. This is particularly clear in Fig. \ref{2danderson0}, for $w\ge 1$; for smaller $w$ the deviation from the $w=0$ behavior and tendency toward saturation is also very obvious. In this case $\ell$ is not well-defined perturbatively due to van Hove singularity, but the localization length $\xi$ is much bigger than the system sizes studied here (numerical calculations of Ref. [\onlinecite{MacKinnonKramer}] show that $\xi$ is in order of $10^4$ for $w=3$ and in order of $10^6$ for $w=2$, and {\em exponentially} bigger for smaller $w$'s), indicating $\xi$ plays no role in the size dependence of EE.

For $E_F=1$ (Fig. \ref{2danderson1}), we again find deviation from the $w=0$ behavior, and tendency toward saturation starts when system size $L$ reaches the mean free path $\ell$. We note in this case we have $\xi\sim\ell e^{\pi k_F\ell/2}\gg\ell$ for small $w$; for example at $w=1$ we expect $\xi\sim 10^{36} \ell$!. We thus again find that while $\xi$ controls the extensiveness of the fermion wave function at the Fermi level, it does {\em not} control the size dependence of EE.

\subsection{Anderson Model in Three Dimensions}

\begin{figure}
\includegraphics[scale=0.2]{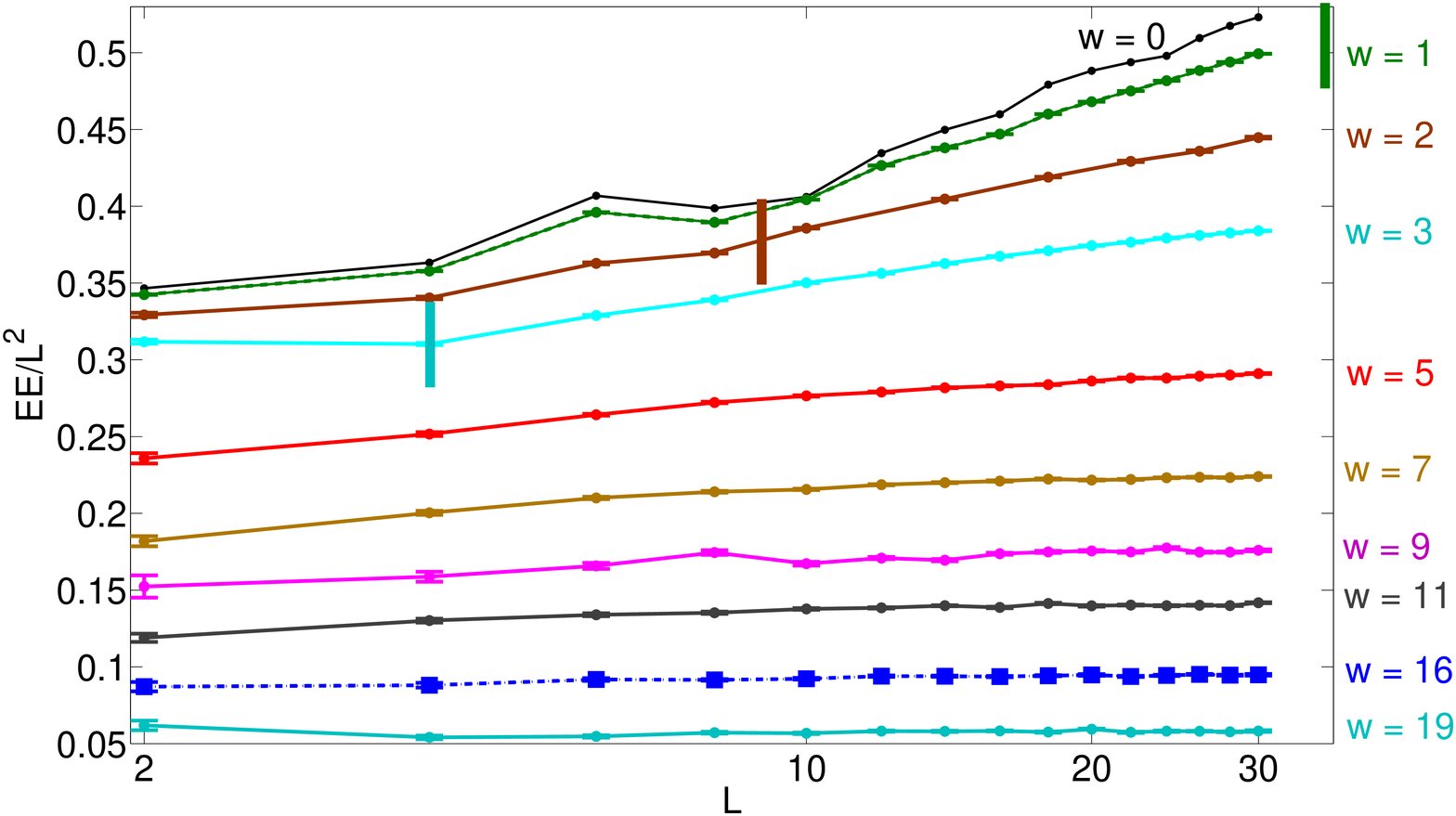}
\caption{\label{3danderson}[Color online] Entanglement Entropy divided by square of linear system size of three dimensional Anderson model in log-linear scale for different value of disorder strength $w$. $E_F=0$. Mean free path corresponding to selected $w$'s is indicated as a vertical line. Data points corresponding to $w_c=16$ are indicated by \textcolor{blue}{$\blacksquare$}. Horizontal axis is the linear size of the system, $L$. The total system has $L \times L \times L$ sites. Number of samples is $100$ for small sizes and $10$ for large sizes.}
\end{figure}

In three dimensions, there is a critical value of disorder strength, $w_c\approx 16$ where a metal-insulator transition occurs. States at the Fermi level are delocalized for $w<w_c$.  In Fig. \ref{3danderson}, EE divided by boundary area $L^2$ (to account for the area law contribution) as a function of $L$ for different values of disorder strength is plotted. Similar to the 1D and 2D cases, we find a logarithmic growth for $w=0$, while for $w > 0$ deviation from such growth, and tendency toward saturation starts when the system size $L$ reaches mean free path $\ell$. It is particularly worth noting that there is {\em no} obvious change of the behavior of EE near $w=w_c\approx 16$, which is highlighted in the figure; area law behavior is clearly seen on both sides of $w_c$. We thus conclude that entanglement area law is respected in both the metallic and insulating phases of the 3D Anderson model, as long as there is finite disorder strength.

\section{Discussion and Summary}

In this work we find, through explicit numerical calculations, that entanglement entropy of free fermion systems obey the area law in the ground states of one-, two- and three-dimensional Anderson models, for subsystems whose linear size is sufficiently bigger than the mean free path. This result holds in the metallic phase of the three-dimensional case, where states at the Fermi energy are extended. Localization length, either finite or infinite, appears to play no special role in the size dependence of entanglement entropy.

This behavior is in sharp contrast to those of two special 1D models studied earlier,\cite{PouranvariYang14} where we found {\em violation} of area law in the metallic phase. We now discuss the source of difference in these models. The first one is the random dimer model,\cite{dunalp} which has a very special feature that there is no back scattering at a special resonance energy, as a result of which the system is metallic when the Fermi energy matches the resonance energy. In this case the mean free path $\ell$ is {\em infinity} (and the fermion motion is ballistic instead of diffusive as in ordinary diffusive metals), resulting in the area law violation, consistent with findings of the present work.

The second special model studied on Ref. \onlinecite{PouranvariYang14} involves power-law long-range hopping,\cite{mirlin} and the system is metallic when the decay exponent is small enough. In this case the fermion can hop over arbitrarily large distances, thus mean free path is not well defined. More importantly, in the metallic phase fermion motion is {\em super}-diffusive,\cite{mirlin} different from that in 3D Anderson model.

Combining with earlier results with the present ones, we conclude that while violation of entanglement area law does not necessarily require a sharp Fermi surface (which is destroyed by disorder), it does not occur in diffusive metals. It can occur, however, in metallic phases of disordered free fermion models where the fermion motion is super-diffusive.

While in this paper we studied non-interacting fermions, a recent paper studied the role of localization length in the case of interacting fermions.\cite{Berkovits} Also another paper\cite{pasturslavin} proved the area law in the localized regime of the Anderson model, although our numerical calculations demonstrate area law in the the metallic phase as well.

{\em Note Added} -- While the present manuscript is being written up, a related preprint\cite{potter} appeared on the arxiv reaching a very similar conclusion. The numerics of that work is limited to quasi-1D strips.

\acknowledgments

This research is supported by DOE grant No. DE-SC0002140.

\appendix

\section{Calculation of Mean Free Path as a Function of Random Potential Strength}
We explain the calculation of mean free path here. We assume that $\varepsilon$'s are uncorrelated random numbers with uniform distribution, i.e. a white noise uniform distribution. First we need to calculate the ensemble averaged transition rate using Fermi's golden rule:

\begin{equation}
\frac{1}{\tau_{\vec{k}}} = \frac{2 \pi}{\hbar} \frac{L^D}{(2 \pi)^D} \int d^D \vec{q}
\avg{ | \langle\psi_{\vec{q}+\vec{k}}
 |\varepsilon_{i}
 \ket{\psi_{\vec{k}}} | ^2}
 \delta (E_{\vec{k}+\vec{q}}-E_{\vec{k}}),
\end{equation}
in which $\avg{}$ stands for ensemble average and $D$ is the dimension. We calculate the mean free path at the Fermi level, which is chosen to be  $E_{F}=0$ for 1D and 3D and $E_{F}=1$ for 2D . Also we know:

\begin{equation}
\avg{ | \bra{\psi_{\vec{q}+\vec{k}}}
 \varepsilon_{i}
 \ket{\psi_{\vec{q}}} | ^2} =
 \sum_{\vec{r}} \sum_{\vec{r}'} \frac{e^{-i \vec{q}.\vec{r}}}{L^D} \frac{e^{+i \vec{q}.\vec{r}'}}{L^D}
\avg{\varepsilon_{\vec{r}} \varepsilon_{\vec{r}'}},
\end{equation}
where for a uniform distribution we have:
\begin{equation}
\avg{\varepsilon_{\vec{r}} \varepsilon_{\vec{r}'}} = \frac{w^2}{12} \delta_{\vec{r},\vec{r}'}.
\end{equation}

Thus, finally we have:

\begin{equation}
\frac{1}{\tau} = \frac{2 \pi}{\hbar} \frac{1}{(2 \pi)^D} \frac{w^2}{12}\int d^D \vec{q}  \\  \delta{(E_{\vec{q}}-E_F)}.
\end{equation}

Then, $\ell = v \tau$, where $v$ is the average velocity of electron on Fermi surface. Calculated mean free path in 1, 2, and 3 dimensions at specific Fermi energy, $E_F$ is:

\begin{equation}
\ell \approx
\left\{
	\begin{array}{ll}
		48/w^2  &E_F =0, D=1  \\
		31/w^2 &E_F=1, D=2  \\
		36/w^2 & E_F=0, D=3
	\end{array}
\right.
\end{equation}

\end{document}